# Role of microenvironment in the mixed Langmuir-Blodgett films.

S. A. Hussain, P. K. Paul and D. Bhattacharjee *

Department of Physics, Tripura University

Suryamaninagar-799130, Tripura, INDIA

**Abstract:**

This paper reports the $\pi$-A isotherms and spectroscopic characteristics of mixed Langmuir and Langmuir-Blodgett (LB) films of non-amphiphilic carbazole (CA) molecules mixed with polymethyl methacrylate (PMMA) and stearic acid (SA). $\pi$-A isotherm studies of mixed monolayer and as well as also the collapse pressure study of isotherms definitely conclude that CA is incorporated into PMMA and SA matrices. However CA is stacked in the PMMA/SA chains and forms microcrystalline aggregates as is evidenced from the scanning electron micrograph picture. Nature of these aggregated species in the mixed LB films has been revealed by UV-Vis absorption and fluorescence spectroscopic studies. The presence of two different kinds of band systems in the fluorescence spectra of the mixed LB films have been observed. This may be due to the formation of low dimensional aggregates in the mixed LB films. Intensity distribution of different band system is highly sensitive to the microenvironment of two different matrices as well as also on the film thickness.

**Key words:**  Nonamphiphilic, Langmuir-Blodgett Films, UV-Vis absorption spectroscopy, Fluorescence spectroscopy.

* Corresponding author
E.mail: tuphysic@sancharnet.in
       debu_bhat@hotmail.com
Phone: +913812375317
Fax: +91-381-2374801

**Introduction:**

Carbazole (CA) and its derivatives have found extensive industrial and commercial applications in the manufacture of organic photoconductors [1] and electroluminescent devices [2-4]. CA and its derivatives are also of great interest to the spectroscopists owing to their unique photo-physical characteristics like an intense and well characterized absorption and fluorescence spectra [5-6]. Owing to their excellent optical and electronic properties, various amphiphilic derivatives of carbazole were synthesized and studied in the Langmuir-Blodgett films [5-13]. However the synthesis of these amphiphilic carbazole derivatives and their purification are extremely difficult. In recent times it was observed that various nonamphiphilic materials could also form well organized LB films when mixed with a long chain fatty acid or inert polymer matrix [14-16]. A preliminary work on the photophysical characteristics of nonamphiphilic carbazole in LB films was reported [6]. In that work aggregation induced reabsorption effect was observed. However detailed investigations of the formation of stable LB films of carbazole doped into various matrices were not reported. Moreover the study of the changes in photophysical properties by changing various LB parameters are also important from the point of view of various technical applications. In this communication, detail investigations on the dependence of photophysical characteristics of mixed LB films of nonamphiphilic carbazole, on various LB parameters, was reported. It was observed that the longer wave length vibrational pattern of fluorescence band system were sensitive to the molefractions, surface pressure of lifting as well as also on the number of layers.

**Experimental:-**

CA was purchased from Aldrich chemical company, USA and vacuum sublimed followed by repeated recrystalization before use. SA (purity > 99%) from Sigma, USA and isotactic PMMA from Polyscience were used as received. The solvent chloroform (SRL, India) was of spectral grade and its fluorescence spectrum was checked before use. A commercially available Langmuir Blodgett film deposition instrument (Apex, 2000C, India) was used for isotherm measurements and for multilayer film deposition. The subphase used was triple distilled deionised water. The pH of the sub phase was 6.5. Solutions of CA, PMMA, SA as well as CA-PMMA and CA-SA mixtures at different molefractions were prepared in chloroform solvent and were spread on the sub phase by a micro-syringe. Surface pressure was recorded by using Wilhelmy plate arrangement as described elsewhere [17]. All isotherms were run several times with freshly prepared solutions. Deposition of multilayers was achieved by allowing the substrate to dip with a speed of 5 mm/min with a drying time of 15 minute after each lift. Fluorescence grade quartz slides were used for spectroscopic measurement. For each molefraction of CA, 10 bi-layer LB films were deposited. We chose 15 mN/m as the standard surface pressure for lifting the LB films for both the matrices. The transfer ratio was found to be $0.98 \pm 0.02$.

Fluorescence spectra and UV-vis absorption spectra were measured by a Perkin Elmer LS 55 spectrophotometer and Perkin Elmer Lambda 25 spectrophotometer respectively. All the measurements were performed at room temperature ($24^O$C). Scanning electron micrographs were taken in a Hitachi (Japan) scanning electron microscope (model S-415A).

**Monolayer characteristics of CA at the air-water interface:**

To study the behaviour of pure CA at the air-water interface, 100 $\mu$l of chloroform solution of CA ($2\times10^{-3}$ M) was spread on the water surface. After allowing the solvent to evaporate for 20 minutes, the barrier was compressed very slowly at a speed of $2\times10^{-3}$ nm$^2$mol$^{-1}$s$^{-1}$. It was observed that the surface pressure rose up to 8 mN/m and the area per molecule was found to be very small. Addition of larger amount of CA solution resulted in the formation of crystalline domains at the air-water interface upon compression, which were visible to the naked eye. Moreover, the islets once formed as a result of barrier compression did not degenerate at the molecular level but remained as smaller islets upon relaxation of surface pressure. Repeated attempts to transfer this floating layer onto solid substrate were failed. However, when CA was mixed with a supporting matrix of either SA or PMMA, the floating layer was found to be highly stable and could be transferred onto quartz substrate.

Figures 1a and 1b show the surface pressure ($\pi$) vs. area per molecule (A) isotherms of CA mixed with PMMA and SA respectively at different molefractions of CA along with pure PMMA and SA. The area per molecule of pure PMMA and SA are 0.11 and 0.23 nm$^2$, respectively at a surface pressure of 15 mN/m, which are consistent with the reported result [14].

Pure PMMA isotherm shows an inflection point at about 20 mN/m [18]. It is also observed that this inflection point gradually loses its distinction with increasing molefractions of CA in PMMA and at and above 0.7M of CA, the CA-PMMA mixed isotherms show steep rising without any transition point. At about or above this inflection point, the collapsing of monolayer occurs and the monolayer remains no longer stable [18]. This is why a surface pressure of 15 mN/m has been chosen for the film deposition.

The most interesting feature of CA-PMMA mixed isotherm is that at lower molefractions of CA upto 0.3 M, the area per molecule of mixed film is larger than that of pure PMMA. Although it gradually decreases with increasing molefractions and from 0.4 M and above molefractions of CA, the area per molecule is lower than that of PMMA. The most plausible explanation seems to be that at lower molefraction of CA, the PMMA matrix behaves as an excellent microenvironment, where CA molecules can accommodate. Thus a monolayer of CA-PMMA is formed. However with increasing molefraction of CA, phase separation between CA and PMMA moieties occur that may originate from the immiscibility of the components owing to their differences in molecular structure, physical and chemical properties, which generates aggregates. In case of isotherm of CA-SA mixed films (figure 1b) also the area per molecule decreases consistently with increasing molefraction.

However negligible small area per molecule of pure CA and at higher molefraction of CA of mixed monolayer, may lead to the conclusion that CA molecules may form aggregates and may be sandwiched between PMMA or SA chains and squeezed into the matrices and out on the air-water interface. Another possibility may be the loss of CA molecules through precipitation into the water subphase.

To confirm whether or not the CA molecules at the air-water interface were lost through precipitation in the bulk of the subphase, small amount of water from just below the air-water interface were sucked out and the fluorescence of the water sample was recorded. It was confirmed from the failure to detect any fluorescence that CA molecules were not lost through submerging below the air-water interface.

Therefore the most possible explanation may be that they are very likely pushed up in between the SA or PMMA chains so as not to occupy any area at the air-water interface.

Moreover an increase in the absorption and fluorescence intensities with increasing number of layers and molefractions of CA in the mixed film (figure not shown) definitely confirm the incorporation of CA molecules into the mixed LB films.

The plot of collapse pressure versus molefraction of CA in the mixed films of CA with PMMA and SA are also shown in the insets of figure 1a and 1b respectively. It is evident from the figures that collapsing of mixed monolayer occurs at much higher surface pressure than that of pure PMMA and SA. This certainly gives the evidence that CA molecules are definitely stacked between PMMA and SA chains and do not precipitate out. Moreover the collapse pressures at various molefractions do not match with the ideality (solid) curve. Which indicate that CA molecules and PMMA/SA molecules are totally immiscible in the mixed monolayer [6]. This immiscibility or complete demixing between CA and PMMA or SA may lead to the formation of crystalline aggregates of CA molecules in the mixed LB films, which has been confirmed later by Scanning electron micrograph (SEM). The nature of these aggregates has been revealed by UV-Vis absorption and fluorescence spectroscopic studies.

**Scanning electron microscopy:**

To visualize the phase separation between unlike molecules and to confirm the formation of microcrystalline aggregates of CA molecules in mixed LB films, a traditional imaging method namely scanning electron micrograph (SEM) has been employed. Figure 2 shows a scanning electron micrograph (SEM) of the 10 layer mixed LB film of 0.5 M of CA in SA matrix. The aggregates with sharp and distinct edges correspond to the three dimensional aggregates of CA in LB films. The smooth background corresponds to the SA matrix. In case of CA-PMMA mixed LB films (figure not shown), crystalline aggregates of CA are also observed. The formation of distinct crystalline domains of CA, as evidenced from the SEM, provides compelling visual evidence of aggregation of CA in the LB films

**UV-Vis absorption and fluorescence spectroscopic study:**

Figures 3a and 3b show the UV-Vis absorption and fluorescence spectra of mixed LB films of CA (0.1-0.8 M) in PMMA and SA matrices respectively along with the spectra in chloroform solution and CA microcrystal for comparison.

Absorption spectrum of CA in chloroform, shows distinct bands within 225-350 nm regions, with intense and sharp high-energy bands having peaks at 247 and 292 nm along with a weak hump at around 258 nm, owing due to $S_2 \rightarrow S_0$ ($^1L_a$-$^1A$) transition [6] that is directed parallel to the long axis of the molecule. There are two other low intense but prominent vibrational bands in the 300-350 nm regions, with maxima at around 319 and 331 nm corresponding to $S_1$-$S_0$ ($^1L_b$-$^1A$) transition [6] that is directed parallel to the short axis of the molecule. The vibrational band system in the CA microcrystal spectrum is almost identical in position and shape except an increase in intensity. However the high-energy bands in microcrystal spectrum become broadened and blue shifted. Also the band with peak at 292 nm in CA solution spectrum is totally absent in microcrystal spectrum.

It is interesting to note that the absorption spectra of the mixed LB films of CA in PMMA/SA matrices, for almost all the molefractions are observed to be diffused, broadened and less intense compared to the solution absorption spectrum except the absorption spectra of mixed LB film of 0.8 M of CA in PMMA. Also the vibrational bands are too weak to be distinguished. However the mixed LB film absorption spectrum of 0.8 M of CA in PMMA matrix becomes almost identical to that of CA microcrystal spectrum. The broadening of the band system may arise due to aggregation. The close similarity of the absorption spectra of the mixed LB films at higher molefractions with that of the microcrystal spectrum may be due to the formation of low dimensional microcrystalline aggregates in the mixed LB films.

It may be mentioned in this context that there are certain rigid nearly planar plate like molecules namely, pyrene [19] or rod like molecules anthracene [20], biquinoline [14] form aggregate in the mixed LB films. These molecules are sandwiched among the matrix molecules (SA/PMMA) to form aggregates in the LB films and partial or total binary demixing is occurred in the mixed LB films. In the present work, CA molecule has almost linear rod like structure. For aggregation to be occurred in the mixed LB films of CA molecules, the most possibility is the sandwich of CA molecules among SA or PMMA molecules in the mixed LB films.

Fluorescence spectrum in chloroform solution (also shown in figures 3a and 3b) shows distinct and prominent vibrational band profile within 325-400 nm region with 0-0 band at 338 nm and the other prominent, intense band at 352 nm. There is a shift of the 0-0 bands at about 7 nm in fluorescence spectrum in comparison to solution absorption spectrum. This may be due to slight deformation of electronic states of the molecules.

The fluorescence spectrum of CA microcrystal is totally different than that of CA solution spectrum. The high energy band system in solution spectrum is totally absent in CA microcrystal and a longer wavelength band system with intense vibrational peaks at 393 and 413 nm along with a weak hump at 436 nm are observed. The origin of this longer wavelength band system is not readily explicable. One possible explanation is that due to closer association of molecules in CA microcrystal, deformation in the electronic structure is occurred resulting in the formation of favourable conditions for transition to the upper vibrational levels of the ground electronic state.

Almost all the bands of CA solution and microcrystal spectra are present in fluorescence spectra of LB films of CA in both the PMMA and SA matrices at different molefractions of CA having 0-0 band at 341 nm, which appears to be slightly shifted with respect to the solution fluorescence spectrum. In comparison to the CA solution spectrum, new bands are observed within 350-450 nm region in the mixed LB films of CA in both the matrices having prominent peaks at 397 and 416 nm along with a weak hump at around 437 nm. It is interesting to note that in case of PMMA matrix with increasing molefraction, the band systems in the 325-375 nm region decrease in intensity and at a molefraction of 0.8 M the band systems totally disappear. However in case of SA matrix the band system within 325-375 nm region become broadened with peak at around 346 nm and remain present in all the molefractions of CA in SA. The decrease in intensity of high-energy band system of 325-375 nm region with respect to the longer wavelength band system in the 350-450 nm region of mixed LB films spectra is an indication of strong aggregation induced reabsorption effect [6] owing to the formation of micro-crystalline aggregates in the mixed LB films.

**Layer effect:**

Various technical applications may sometimes require thick films. Here we have also studied the dependence of photophysical characteristics on the number of layers. Figures 4a and 4b show the UV-Vis absorption and steady state fluorescence spectra of different layered mixed LB films of CA in PMMA and SA matrices respectively, at two different molefractions of 0.1 M and 0.5 M of CA.

No appreciable change in band position is observed in the UV-Vis absorption spectra of the mixed LB films in both the matrices except a little change in intensity distribution. However their fluorescence spectra are quite interesting. The fluorescence spectra of the mixed LB films of CA in PMMA matrix (figure 4a) show that when the number of layer is small, namely five, then at both the molefractions of 0.1M and 0.5M, the longer wavelength band system with prominent vibrational peaks at 397, 416 and 436 nm are quite intense in comparison to the high energy band. The high-energy band system is quite low intense having overlapping of several diffused vibrational peaks. However 25 layered LB films for both the molefractions show intense high energy bands with prominent vibrational peaks at 342 and 353 nm and quite low intense longer wavelength band system. However in case of SA matrix, even in case 25 layer LB films, both the longer and shorter wavelength band systems remain stronger.

The increase in high energy band in case of PMMA matrix at higher number of layers may be due to the fact that some conformational changes in CA molecules occur in the microenvironment of PMMA matrix. As a result, deformation in electronic structure tends to reduce. Moreover a comparison of the two different matrices may definitely lead to the conclusion that the conformational and organizational changes of CA molecules are largely affected by the different kinds of microenvironment in two different matrices.

**Pressure effect:**

By changing the surface pressure of lifting, the morphology and crystal parameters of the LB films may be controlled precisely. Figure 5 shows the spectroscopic characterizations of the mixed LB films of CA in SA matrix at two different molefractions (0.1 and 0.5M) of CA using 15, 20, 25 and 30 mN/m surface pressure. All the films are of 10 layers thick.

No appreciable change in fluorescence spectra is observed with change in pressure. Only at higher surface pressure of 30 mN/m and also at higher molefraction of 0.5 M the longer wavelength band systems become intense in comparison to the high energy band. A comparison with the microcrystal fluorescence spectrum of CA shows that at higher surface pressure, the CA molecules in the mixed LB films organized somewhat in a large dimensional microcrystal form.

**Conclusion**:

In conclusion, our results show that carbazole molecules form an excellent Langmuir monolayer at the air-water interface when incorporated into PMMA or SA matrices and can be easily transferred onto solid substrate to form mono- and multilayered Langmuir-Blodgett (LB) films. Isotherms as well as molefraction versus area per molecule studies reveal that in case of PMMA matrix a phase separation occurs between CA and PMMA molecules, which leads to the formation clusters or microcrystalline aggregates of CA molecules in PMMA mixed LB films. However, in case of SA matrix, from the above studies, we cannot say definitely whether phase separation occurs in SA mixed LB films. The collapse pressure versus molefraction plot of CA in SA mixed films clearly reveals the nature of complete demixing of the binary components in the CA-SA mixed Langmuir films. Scanning electron micrograph gives visual evidence of microcrystalline aggregates of CA molecules in the mixed LB films. UV-Vis absorption and steady state fluorescence spectroscopic studies definitely conclude that CA molecules form microcrystalline aggregates in the mixed LB films in case of both the matrices. Fluorescence spectra of mixed LB films of CA in both the matrices of PMMA and SA at different molefractions reveal two different band systems in the longer wavelength region and in the shorter wavelength or high-energy region. Fluorescence spectroscopic study of different number layered mixed LB films in two different matrices reveal that microenvironment plays an important role in conformational and organizational changes of CA molecules in the mixed LB films.

**Acknowledgement:**

The authors are grateful to DST, Govt. of India for providing financial assistance through SERC-DST project No. SR/FTP/PS-05/2001.

**Figure Captions**

Fig: 1(a): Surface pressure ($\pi$) versus area per molecule (A) isotherms of CA in PMMA matrix at different molefractions of CA. The numbers denote corresponding molefractions of CA in PMMA matrix. PM and CA correspond to the pure PMMA and pure CA isotherms. The inset shows the plot of collapse pressures with molefractions of CA in PMMA matrix. Molecular structure of CA is also shown in the inset of 1a.

Fig: 1(b): Surface pressure ($\pi$) versus area per molecule (A) isotherms of CA in SA matrix at different mole fractions of CA. The numbers denote corresponding molefractions of CA in SA matrix. SA and CA correspond to the pure SA and pure CA isotherms. The inset shows the plot of collapse pressures with molefractions of CA in SA matrix.

Fig: 2: Scanning Electron Micrograph (SEM) of 10 layers of CA-SA mixed LB films (molefraction of CA = 0.5 M) at room temperature.

Fig: 3(a): UV-Vis absorption and fluorescence spectra of CA in chloroform solution ($CHCl_3$), in microcrystal (MC) and in CA/PMMA mixed LB films. The numbers denote the corresponding molefractions of CA in PMMA matrix.

Fig: 3(b): UV-Vis absorption and steady state fluorescence spectra of CA in chloroform solution ($CHCl_3$), in microcrystal (MC) and in CA/SA mixed LB films. The numbers denote the corresponding molefractions of CA in SA matrix.

Fig: 4(a): UV-Vis absorption and steady state fluorescence spectra of different layer mixed LB films of CA in PMMA matrix at two different molefractions of 0.1 and 0.5 M of CA. The numbers denote the corresponding layer and molefraction.

Fig: 4(b): UV-Vis absorption and steady state fluorescence spectra of different layer mixed LB films of CA in SA matrix at two different molefractions of 0.1 and 0.5 M of CA. The numbers denote the corresponding layer and molefraction.

Fig: 5: Fluorescence spectra of CA/SA mixed LB films at different surface pressures of lifting at two different molefractions of 0.1 and 0.5 M of CA. The numbers denote the corresponding surface pressure of lifting and molefraction.

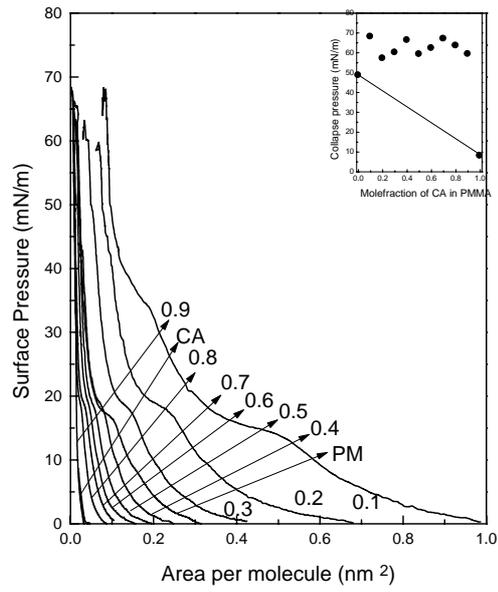

Figure: 1(a) S. A. Hussain et. al

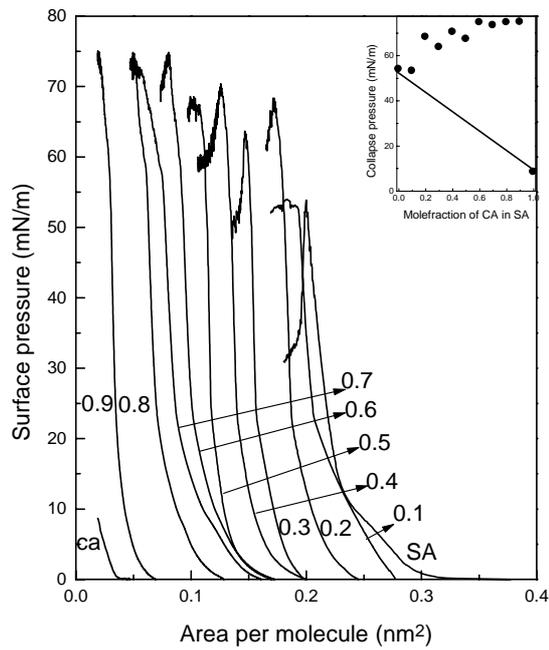

Figure: 1(b) S. A. Hussain et. al

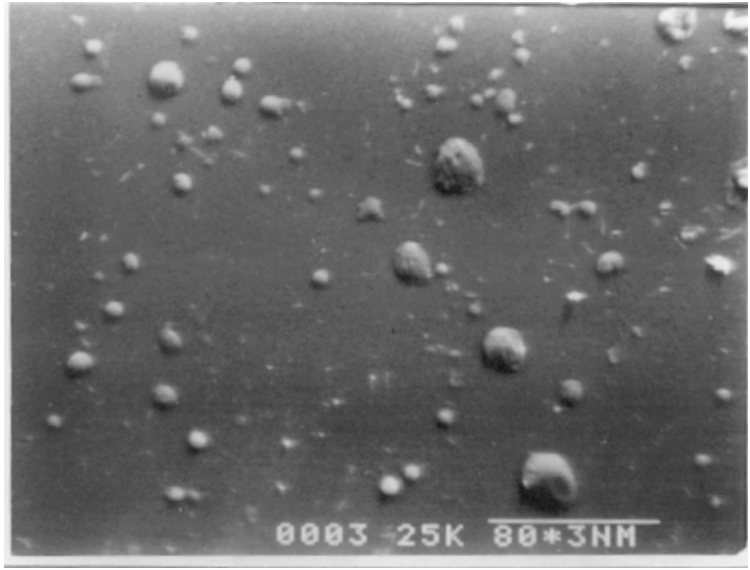

Figure: 2 S. A. Hussain et. al

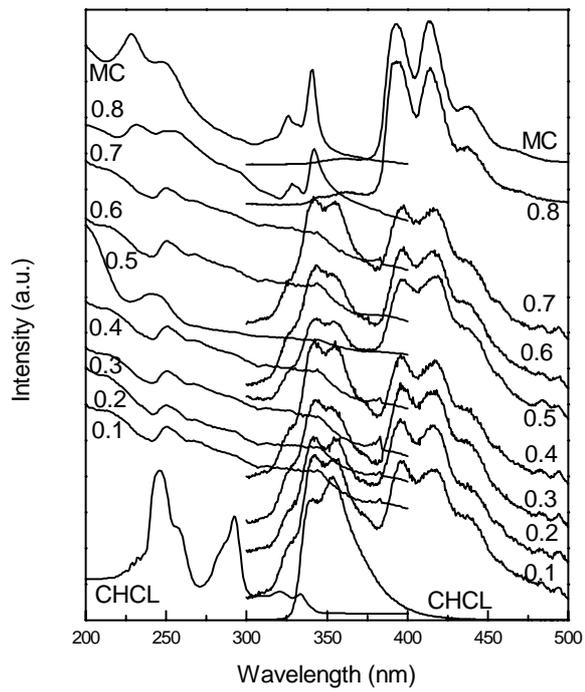

Figure: 3(a) S. A. Hussain et. al

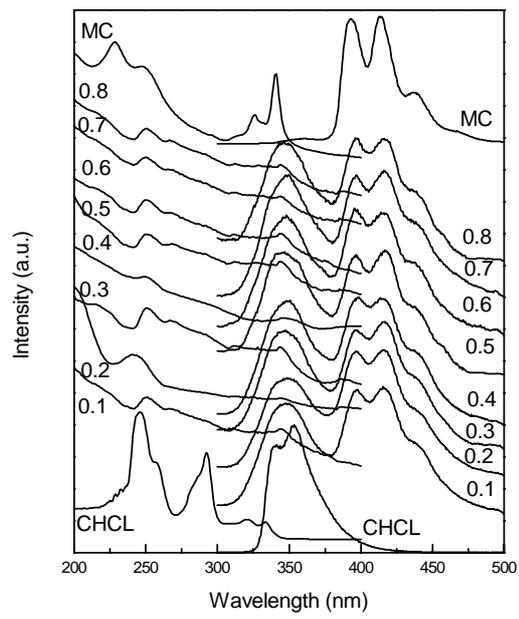

Figure: 3(b) S. A. Hussain et. al

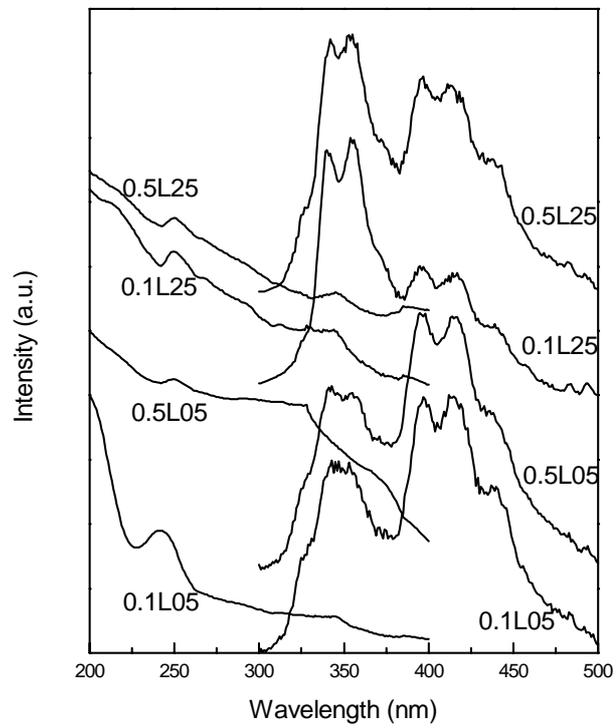

Figure: 4(a) S. A. Hussain et. al

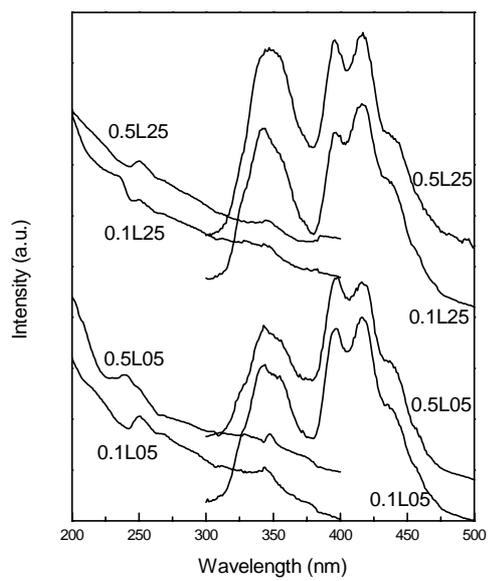

Figure: 4(b) S. A. Hussain et. al

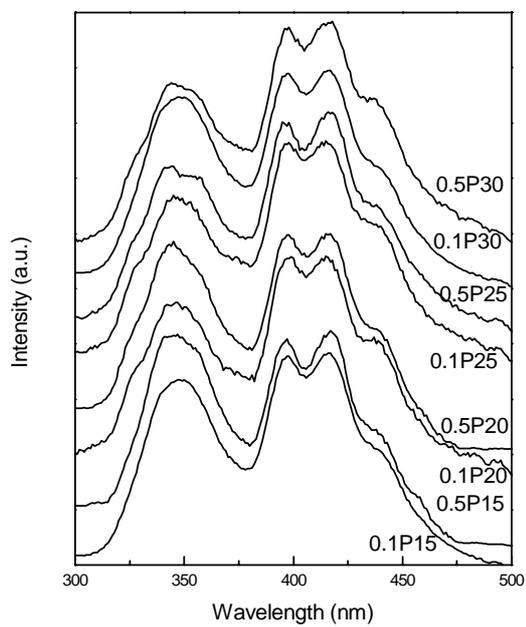

Figure: 5 S. A. Hussain et. al